\documentclass[prb,twocolumn,showpacs,aps,superscriptaddress,floatfix]{revtex4}
\usepackage{amsmath}
\usepackage{amssymb}
\usepackage{bm}
\usepackage{graphicx}
\usepackage{color}
\usepackage{hyperref}
\usepackage{verbatim}
\usepackage[T1]{fontenc}

\begin{document}

\title{Magic radius of AA bilayer graphene quantum dot}

\author{A.L. Rakhmanov}
\affiliation{Institute for Theoretical and Applied Electrodynamics, Russian
Academy of Sciences, 125412 Moscow, Russia}

\author{A.V. Rozhkov}
\affiliation{Institute for Theoretical and Applied Electrodynamics, Russian
Academy of Sciences, 125412 Moscow, Russia}

\author{A.O. Sboychakov}
\affiliation{Institute for Theoretical and Applied Electrodynamics, Russian
Academy of Sciences, 125412 Moscow, Russia}

\begin{abstract}
We study analytically and numerically electronic properties of a circular
quantum dot made from AA bilayer graphene. We observe a discrete set of dot
radii for which the low-energy electron states are degenerate with respect
to the layer parity. By analogy with the ``magic angles" in the twisted
bilayer graphene we refer to these radii as ``magic". Such a feature is
unique for the AA structures and is related to a specific layer-symmetry of
the AA graphene bilayer: the parity of the highest occupied level changes
from layer-symmetric to layer-antisymmetric when the radius of the AA dot
is equal to its magic value. We explore an analogy in the electronic
structure between twisted bilayer graphene at the magic twist angle and the
AA quantum dot with magic radius. We argue that this analogy can be helpful
for theoretical description of the electronic properties of the twisted
bilayer graphene.
\end{abstract}

\pacs{Graphene}

\date{\today}

\maketitle

\section{Introduction}

In the last years we witnessed a great success in fabrication of two-dimensional (2D) van der Waals heterostructures. Current level of technology allows for extraordinary degree of control over the geometry and composition of the artificially created 2D systems. These advances open new research directions and pose original questions for both theorists and experimentalists. Motivated by these factors, in this paper we study an interesting mesoscopic object, a quantum dot cut out from the AA bilayer graphene (AA-BLG). The AA-stacked bilayer graphene is one of the two types of high-symmetry stacking of graphene bilayer (the other is an AB-, or Bernal, type)~\cite{ourBLGreview2016}. Although AA-BLG is less studied experimentally, there are published  reports~\cite{aa_graphit_roy1998,aa_graphite_lee2008, liu_aa_exp2009,
borysiuk2011_aa} on the AA-BLG samples preparation. There is also a rich theoretical research of the AA-stacked graphite/graphene electronics properties~\cite{ourBLGreview2016,aa_tb_charlier1991,aa_abinit_charlier1992,
aa_graphite_charlier1994, aa_graph2012_prl,aa_graph2013,graph_phasep2013,
aa_graph2014,graphIt_aa, brey_fertig_aa2013, fracmet2021prblett,
optical_theor_aa2010, bi3, optical_theor_aa_ab2012, aa_graph_afm2021theor}.

Here, we investigate, both analytically and numerically, electronic properties of a circular AA-BLG quantum dots (AA-QD) disregarding electron-electron interaction. We discover that there is a discrete set of dot radii for which the electron states of the dot are extra degenerate. By analogy with ``magic angles" in the theory of the twisted bilayer graphene (tBLG) we refer to these radii as ``magic".

The magic radius degeneracy originates from the disappearance of the gap
between the lowest empty and the highest occupied single-electron states.
Such a feature is exclusive to the AA-QD and is stable against a change of
the boundary conditions and/or the shape of the AA-QD. It relates to a
specific topological property of the chiral electrons in the AA-BLG: the
symmetry of the wave function for the highest occupied level changes from
layer-symmetric to layer-antisymmetric when radius of the AA-QD is equal to
its magic value. The energy spectra of quantum dots cut out either from the
single-layer (SLG) or from the AB-stacked bilayer graphene (AB-BLG)
demonstrate qualitatively different behavior and do not show this extra
degeneracy since their Hamiltonians do not possess the mentioned above
symmetry.

The identification of the magic radius sequence is the main result of this
paper. Our findings highlight unique properties of AA-QD among other
graphene mesoscopic objects: the degeneracy of the dot ground state allows
for stabilization of interaction-driven exotic ordered states such that the
spin-valley
half-metal~\cite{spin_valley2017prl}
or even ``fractional metal''
phases~\cite{fracmet2021prblett}.
In addition, we expect that our study will be helpful for further
theoretical description of the electronic properties of the tBLG with magic
values of the twist angle. To this end, we will argue that there is a
simple connection between magic twist angles of the tBLG and magic radii of
the AA-QD. This correspondence is based on the fact that a tBLG sample may
be considered as a structure consisting of the AA- and AB-stacked regions,
with the conduction electrons residing mainly in the AA
parts~\cite{ourBLGreview2016}.

The paper is organized as follows. In
Sec.~\ref{Cont_Mod},
to describe electronic properties of the AA-BLG, we outline a continuous
model which is then used for analytical calculations of the energy spectrum
of the AA-QD. The sequence of the calculated magic radii is analyzed in
Sec.~\ref{MagicRad}.
Further,
Sec.~\ref{Numerical}
presents and discusses numerical data for the spectrum of the AA-QD which
are used to generalize and check the analytical results. In
Sec.~\ref{Discussion}
we discuss the magic sizes of the triangular AA-QD, compare the energy
spectrum of the AA-QD with that of similar SLG and AB-BLG quantum dots and,
finally, outline a possible analogy between properties of the AA-QD with
the magic radius and the tBLG with the magic twist angle. Conclusions are
presented in
Sec.~\ref{Conclusions}.

\section{Electron spectrum of an AA-QD}
\label{Cont_Mod}

\subsection{Continuous model for the AA-BLG}

Several theoretical approaches can be used to identify the magic radii
sequence of the AA-QD. In this section we apply long-wave-length, or
continuous, approach and the Dirac equations to calculate energy levels of
the AA-QD. This approximation is suitable since here we are interested only
in the low-energy states of sufficiently large graphene structures. We
introduce index $\xi$ that labels two graphene valleys in the
${\bf k}$-space:
$\xi = 1$ ($\xi=-1$)
represents valley
${\bf K}$
(valley
${\bf K}'$).
Within the valley $\xi$, the single-electron Dirac-like equation for the
AA-BLG reads
\begin{eqnarray}
\label{eq::Dirac_BI_layer}
\varepsilon \psi_{i \xi}
=
t_0 \psi_{\bar{i} \xi} + H_{\rm SLG}^{\xi} \psi_{i \xi}.
\end{eqnarray}
Here
$\psi_{i\xi} = (a_{i\xi}, b_{i\xi})^T$
is a spinor, whose component $a$ (component $b$) represents the wave
function on sublattice $A$ (on sublattice $B$), index
$i=1,2$
labels graphene layers,
$t_0 \approx 0.35$\,eV
is the inter-layer hopping amplitude, and symbol
$\bar{i}$
denotes `not~$i$' layer. The SLG Hamiltonians
$H^{\bf K, K'}_{\rm SLG}$
are the Dirac differential operators
\begin{eqnarray}
\label{DiracEq}
H^{\bf K}_{\rm SLG} = -i\hbar v_{\rm F} \boldsymbol\sigma\cdot\bm{\nabla}
=
\hbar v_{\rm F}\left(
  \begin{array}{cc}
    0 & -i\partial_z \\
    -i\overline\partial_{z} & 0 \\
  \end{array}
\right),
\\
\label{DiracEq1}
H^{\bf K'}_{\rm SLG} = -i\hbar v_{\rm F} \boldsymbol{\sigma}^*\cdot\bm{\nabla}
=
\hbar v_{\rm F}\left(
  \begin{array}{cc}
    0 & -i\overline\partial_{z} \\
    -i\partial_{z} & 0 \\
  \end{array}
\right),
\end{eqnarray}
where $\boldsymbol{\sigma} = (\sigma_x, \sigma_y)$
is the 2D vector of the Pauli matrices acting in the sublattice space,
$\bm{\nabla} = (\partial_x, \partial_y)$
is the 2D gradient operator,
$v_{\rm F}$
is the Fermi velocity, and differential operators
$\partial_z$
and~$\overline\partial_{z}$
are defined as
$\partial_z = \partial_x - i \partial_y$, $\overline\partial_{z} =
\partial_x + i \partial_y$.

Near valley
${\bf K}$,
Eq.~\eqref{eq::Dirac_BI_layer}
can be written as
\begin{eqnarray}
\label{eq::Dirac_dim1}
-i\partial_z b_{i {\bf K}} + a_{{\bar i}{\bf K}} = E a_{i {\bf K}},
\\
\label{eq::Dirac_dim2}
-i\overline \partial_{z} a_{i {\bf K}} + b_{\bar i {\bf K}} = E b_{i {\bf K}}.
\end{eqnarray}
Here the dimensionless eigenenergy is $E = \varepsilon / t_0$, and the distance is measured in terms of a length scale $l$ defined as
\begin{equation}
\label{denota}
l=\frac{\hbar v_{\rm F}}{t_0}=\frac{3a_0t}{2t_0}\sim12a_0\,.
\end{equation}
In this expression the amplitude
$t\approx2.7$\,eV 
describes the in-plane nearest-neighbor hopping, which is related to the
Fermi velocity as
$\hbar v_{\rm F} = 3 t a_0/2$,
symbol
$a_0$
being the shortest C-C distance in graphene. To obtain the Dirac-like
equations for the valley
${\bf K}'$
we should only swap operators 
$\partial_z$
and
$\overline \partial_z$
in
Eqs.~\eqref{eq::Dirac_dim1}
and~\eqref{eq::Dirac_dim2},
see
Eqs.~\eqref{DiracEq}
and~\eqref{DiracEq1}.

Due to the layer-symmetry, it is convenient to introduce $a_{\pm {\bf K}}=a_{1 {\bf K}}\pm a_{2{\bf K}}$, $b_{\pm {\bf K}} =b_{1{\bf K}}\pm b_{2{\bf K}}$, and replace Eqs.~(\ref{eq::Dirac_dim1}) and~(\ref{eq::Dirac_dim2}) by an equivalent set
\begin{eqnarray}
\label{eq::Dirac1}
  i \partial_z b_\pm \mp (1\mp E)a_\pm&=& 0\,, \\
\label{eq::Dirac2}
  - i \overline \partial_z a_\pm \pm (1 \mp E)b_\pm &=& 0\,,
\end{eqnarray}
where we suppress valley index $\xi = {\bf K}$. Since for ${\bf K}'$~valley the derivation is identical, we do not discuss it.

To solve
Eqs.~(\ref{eq::Dirac1})
and~(\ref{eq::Dirac2}),
we exclude
$b_\pm$
from them and derive a Helmholtz-type equation for $a_\pm$
\begin{eqnarray}
\label{eq::helmgoltz_eq}
  \triangle a_\pm +(1 \mp E)^2a_\pm =0\,,
\end{eqnarray}
where
$\triangle=\partial_x^2+\partial_y^2$
is the Laplace operator in the dimensionless coordinate.

The basic object of our study is a sample of AA-BLG in the form of a disc
of radius $R$, see Fig.~\ref{fig::edge_a}. We will show that the magic
radii are much larger than
$a_0$.
Thus, we consider only large dots:
$R\gg a_0$.
To obtain the low-energy properties of such a dot, one must solve
Eq.~(\ref{eq::helmgoltz_eq})
and use the solution to recover all four components of the wave function
$\psi_{1,2}$.
The result reads
\begin{eqnarray}
\label{eq::solutionBLG}
\nonumber
a_{1}&=& e^{i \mu \theta}\left\{C_{\mu}J_{\mu}[(1-E)r]+B_{\mu}J_{\mu}[(1+E)r]\right\},\\
b_{1}&=&-ie^{i(\mu+1)\theta}\left\{C_{\mu}J_{\mu+1}[(1-E)r]\right.\nonumber\\
&&\left.-B_{\mu}J_{\mu+1}[(1+E)r]\right\},\\
\nonumber
a_{2}&=&e^{i \mu \theta}\left\{C_{\mu}J_{\mu}[(1-E)r]-B_{\mu}J_{\mu}[(1+E)r]\right\},\\
b_{2}&=&-ie^{i(\mu+1)\theta}\left\{C_{\mu}J_{\mu+1}[(1-E)r]\right.\nonumber\\
\nonumber
&&\left.+B_{\mu}J_{\mu+1}[(1+E)r]\right\},
\end{eqnarray}
where ($r$, $\theta$) are the polar coordinates, $J_\mu (x)$ are the Bessel functions, $\mu \in \mathbb{Z}$ is the quantized angular momentum of the solution, and $C_\mu$, $B_\mu$ are complex coefficients.

\subsection{Boundary condition}

Radial component
$j_r$
of the current is zero at the dot boundary
$r=R$.
Then, following
Ref.~\onlinecite{akhmerov},
we can formulate a proper boundary condition for the wave
function~(\ref{eq::solutionBLG}).

We are interested in the low-energy spectrum of the dot. Thus, for
sufficiently smooth edges, the intervalley electron scattering can be
neglected. Under this assumption the boundary condition
reads~\cite{akhmerov}
\begin{eqnarray}
\label{eq::dot_BC}
[M_0 (\theta) - I] \psi_i (R, \theta) = 0,
\quad
i = 1,2,
\end{eqnarray}
for any angle $\theta$. Matrix $M_0 (\theta)$ is
\begin{eqnarray}
\label{matrixSLG}
M_0\!=\!\left(
\begin{array}{cc}
\cos \phi & i e^{-i\theta} \sin \phi \\
-i e^{i\theta} \sin \phi & -\cos \phi \\
\end{array}
\right),
\end{eqnarray}
where
$-\pi < \phi < \pi$
is a phenomenological parameter determined by detailed structure of the
edge of the dot. In principle, $\phi$ may vary with $\theta$, but we will
ignore such a possibility in this section. The matrix in
Eq.~\eqref{matrixSLG}
is the most general form of the unitary Hermitian 2$\times$2 matrix
($M_0^{-1} = M_0^\dag = M^{\vphantom{\dagger}}_0$)
that anticommutes with the radial current operator
\begin{eqnarray}
j_{r} = \hbar v_{\rm F} (\sigma_x \cos \theta + \sigma_y \sin \theta)\,.
\end{eqnarray}
Condition~(\ref{eq::dot_BC})
guarantees that the current normal to the dot's edge
$j_\perp = j_r$
vanishes at the edge for both graphene layers individually.

\begin{figure}[t]
\includegraphics[width=0.45\columnwidth]{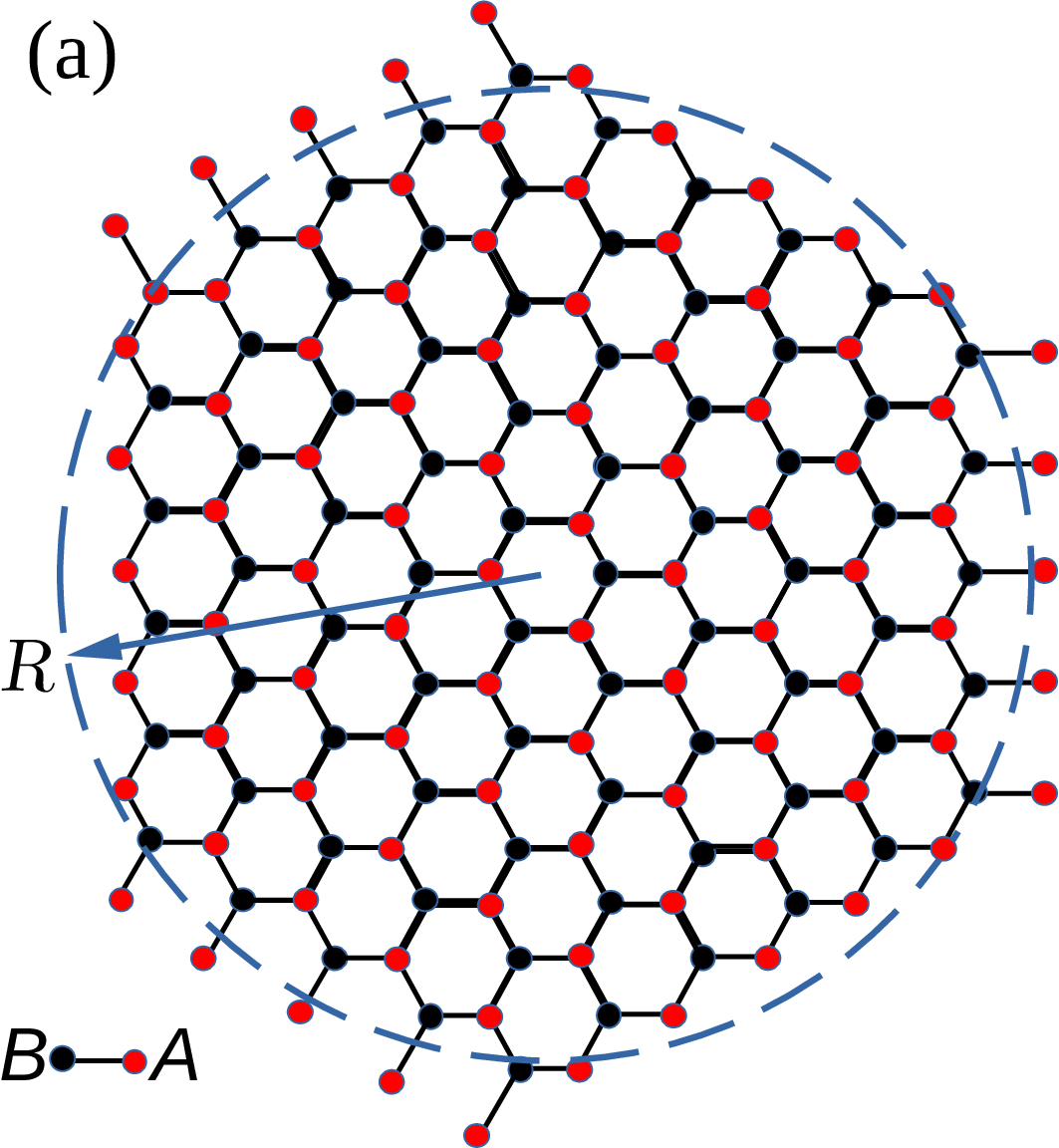}
\qquad
\includegraphics[width=0.45\columnwidth]{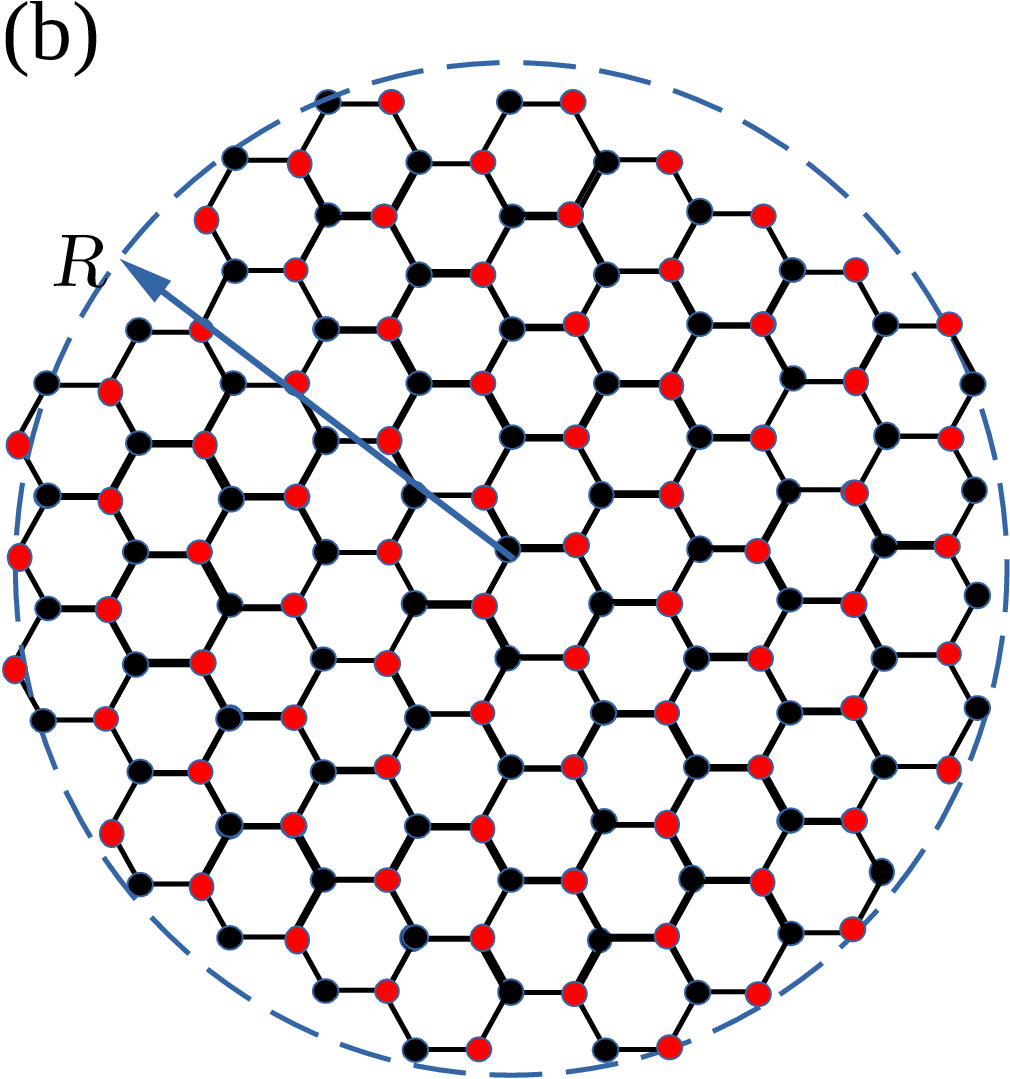}
\caption{AA-BLG quantum dots, top view (lower layer is not visible).
Sublattice $A$ ($B$) is shown by red (black) circles. (a)~A dot whose edge
atoms are entirely in $A$~sublattice. (b)~A circular quantum dot (note
irregular structure of the edge). The dots' shapes can be approximated by
circles of radius $R$ [plotted as (blue) dashed lines].
\label{fig::edge_a}
}
\end{figure}

Equations~\eqref{eq::dot_BC} form a system of four linear equations for
two independent constants $C_\mu$ and $B_\mu$. However, the conditions
imposed on the matrix $M_0$, in particular, mean that
$\textrm{det}(M_0-I)=0$, and the rank of this system is reduced to two. As a result, Eq.~\eqref{eq::dot_BC} allows one to derive unambiguously the energy spectrum of the dot.

For $\phi = 0, \pi$ the boundary
condition~(\ref{eq::dot_BC})
requires that either
$a_i$
or
$b_i$
vanishes at the dot edge. Thus, such a choice of $\phi$ may be used for
approximate description of a dot whose edge atoms all belong to the same
sublattice, as depicted, for example, in
Fig.~\ref{fig::edge_a}\,(a).
When
$\phi = \pm \pi/2$,
we have a type of infinite-mass boundary
condition~\cite{akhmerov}.

\subsection{Quantum dot spectrum}

A straightforward algebra allows to establish that the homogeneous system
of the linear
equations~(\ref{eq::dot_BC})
for $C_\mu$ and $B_\mu$ has a non-zero solution only when the following
conditions are satisfied
\begin{eqnarray}
\label{eq::spectrum}
J_{\mu}[(E - c) R]\!  +\!  \cot\/\! \left( \frac{\phi}{2} \right)
J_{\mu+1}[(E - c) R]=0,
\end{eqnarray}
where $c=\pm 1$. These relations, treated as equations for $E$, determine the single-electron spectrum for a quantum dot of radius $R$.

The eigenvalues with $c=+1$ correspond to the wave function in
Eq.~\eqref{eq::solutionBLG}
with
$B_\mu=0$,
which is equivalent to
$a_1=a_2$ and $b_1=b_2$.
These states will be referred to as layer-symmetric. The eigenvalues with
$c=-1$
correspond to
$C_\mu=0$,
which implies that
$a_1=-a_2$
and
$b_1=-b_2$.
We call these states layer-antisymmetric. The quantum number $c$ is an
eigenvalue of the layer parity
operator~\cite{ourBLGreview2016,PhysRevB.88.245404}
\begin{equation}
\label{C_operator}
\hat{C}=\tau_x\otimes \sigma_0\,,
\end{equation}
which commutes with the
Hamiltonian~\eqref{eq::Dirac_BI_layer}.
Here
$\tau_x$
is the Pauli matrix in the layer space,
$\sigma_0$
is the identity matrix in sublattice space, and $\otimes$ denotes direct
product of the matrices. This symmetry and corresponding conserved quantum
number is a characteristic property of the AA-BLG, which is
absent~\cite{ourBLGreview2016}
for both the SLG and AB-BLG.

\begin{figure}[t]
\includegraphics[width=0.99\columnwidth]{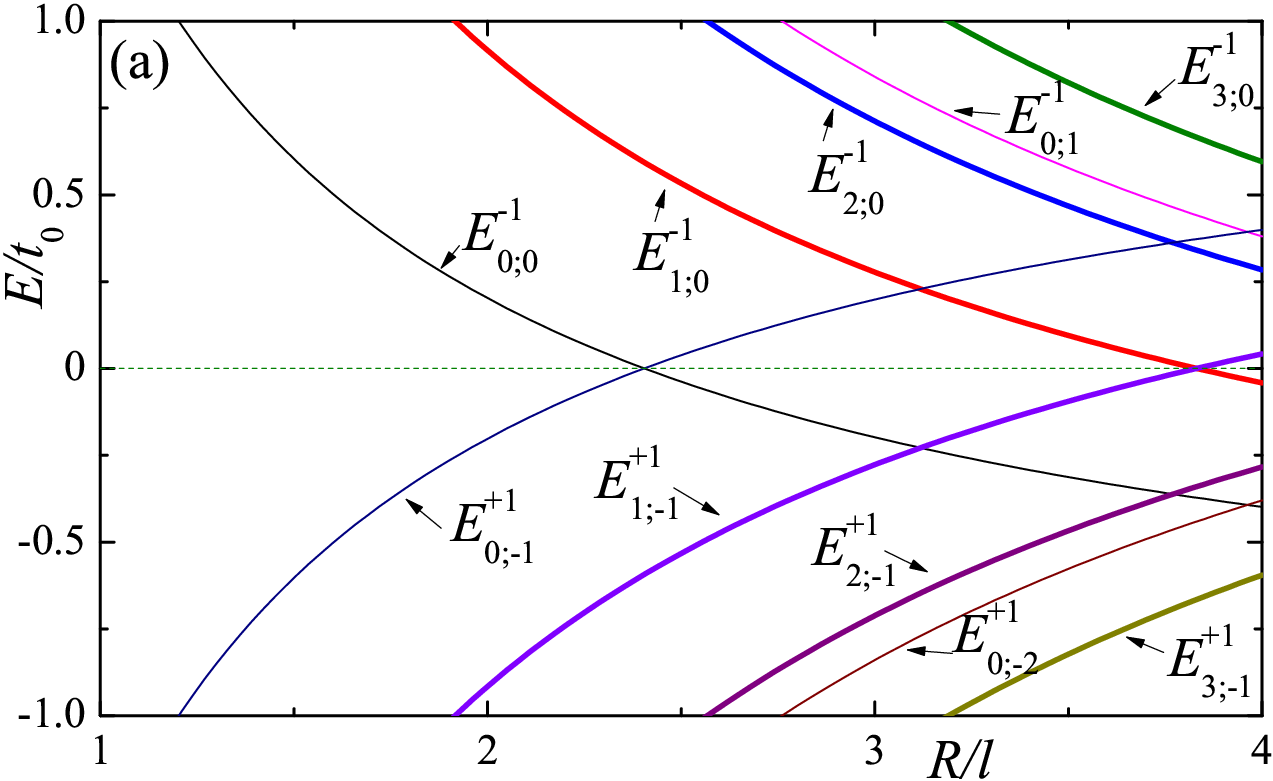}\\
\includegraphics[width=0.99\columnwidth]{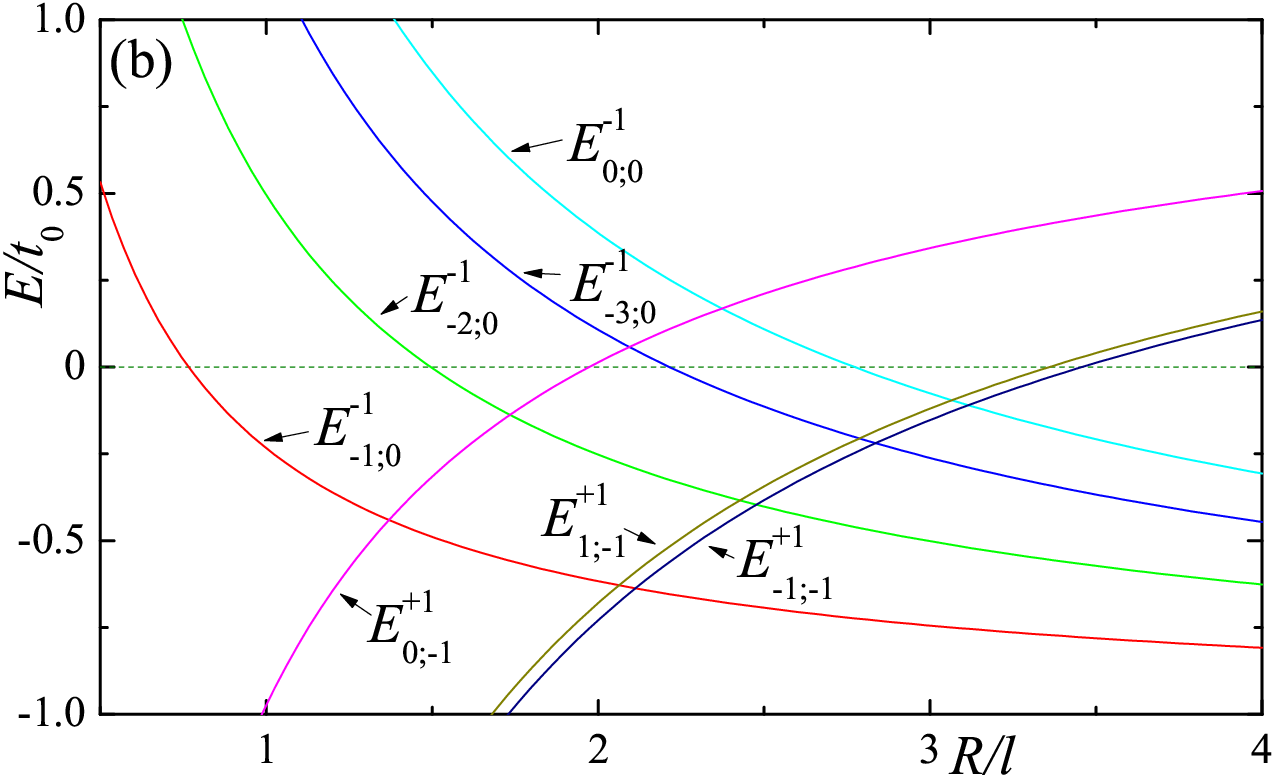}\\
\includegraphics[width=0.99\columnwidth]{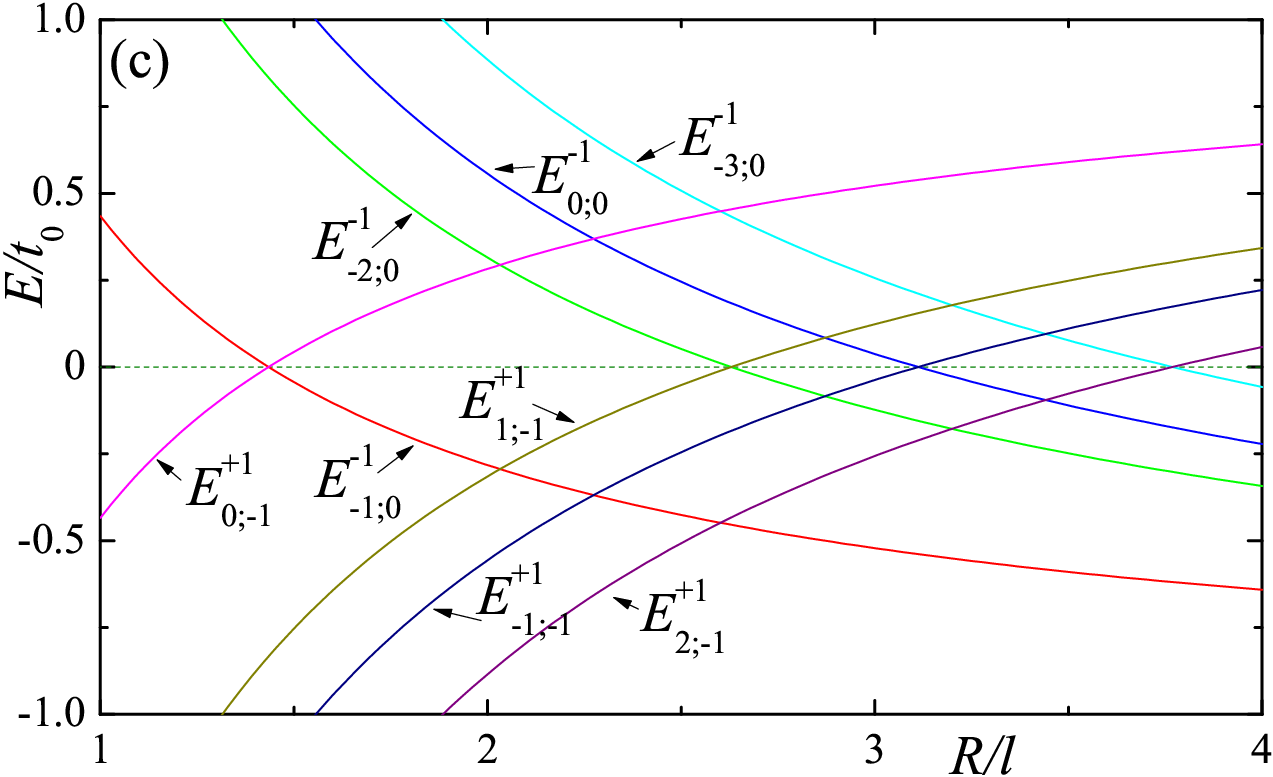}
\caption{First several eigenenergies
$E^{c}_{\mu;n}$
close to zero energy as function of the sample's radius $R$ calculated for
$\phi = \pi$
[panel (a)],
$\phi = 3\pi/4$
[panel (b)], and
$\phi = \pi/2$
[panel (c)]. The eigenenergies shown by thick curves in panel (a) are
doubly degenerate (this is a consequence of
$E^{c}_{\mu;n}$
being independent of the sing of $\mu$ for
$\phi = \pi$).
\label{fig::spectrum}
}
\end{figure}

Multiple solutions of Eq.~(\ref{eq::spectrum}) can be compactly expressed as
\begin{equation}
\label{eq::spectrum_compact}
E_{\mu;n}^{c} (R)
=
c +\frac{z_{\mu;n}}{R},
\end{equation}
where $z_{\mu;n} = z_{\mu;n} (\phi)$ is the $n$-th ($n \in \mathbb{Z}$) root of the equation
\begin{equation}
\label{eq::Bessel0}
J_{\mu}(z)+\cot\!\left(\frac{\phi}{2}\right) J_{\mu+1}(z)=0\,.
\end{equation}
Index $n$ can be viewed as the principle quantum number for an electron
state
$E^{c}_{\mu;n}$.
It enumerates the roots of Eq.~(\ref{eq::Bessel0}) in such a manner that
$z_{\mu;n+1}>z_{\mu;n}$, and $z_{\mu;0}$
is the first positive root in the sequence.
Figure~\ref{fig::spectrum}
shows the dot spectrum for various $R$ and $\phi$.

The AA-QD spectrum demonstrates extra degeneracies when $\phi$ equals zero or $\pi$ [see bold lines in Fig.~\ref{fig::spectrum}(a)]. Indeed, let $\phi = \pi$. In this case Eq.~\eqref{eq::Bessel0} reduces to $J_\mu (z) = 0$. Using $j_{\mu;n}$ as a notation for the $n$-th positive root of the Bessel function $J_\mu$, we can write
\begin{eqnarray}
\label{eq::z_eq_j_a}
&& z_{\mu;n} (\pi) = j_{\mu;n+1}, {\rm \ when}\quad n \geq 0\,,
\\
\label{eq::z_eq_j_b}
&& z_{\mu; n} (\pi) = - j_{\mu; -n}, {\rm \ when}\quad n < 0\,.
\end{eqnarray}
Since $j_{\mu; n}$ is an even function of $\mu$, we have $E^{c}_{\mu; n} =
E^{c}_{-\mu; n}$ for $\phi = \pi$. Thus, if $\phi=\pi$, all energy levels
$E^{c}_{\mu;n}$ with $\mu\neq 0$ are double-degenerate (in addition to spin
and valley degeneracy). Similar reasoning for the case $\phi=0$ proves that
$E^{c}_{\mu; n} = E^{c}_{-\mu-2; n}$,
and the eigenstates are double-degenerate if
$\mu\neq -1$.

Note also that Eqs.~(\ref{eq::spectrum_compact}), (\ref{eq::z_eq_j_a}), and (\ref{eq::z_eq_j_b}) imply that at $\phi = 0,\, \pi$ the eigenvalues satisfy a relation
\begin{eqnarray}
\label{eq::spectral_symm}
E_{\mu; n}^{c} = - E_{\mu;-n-1}^{-c},\quad n \geq 0\,.
\end{eqnarray}
This property can be traced back to the chiral (sublattice) symmetry of the
AA-BLG Hamiltonian with nearest-neighbor hopping. Indeed, one can check
that the transformation
\begin{eqnarray}
\label{eq::sublattice_symm}
a_i \rightarrow (-1)^i a_i,
\quad
b_i \rightarrow (-1)^{i+1} b_i
\end{eqnarray}
changes the sign of the eigenvalue in Eq.~(\ref{eq::Dirac_BI_layer}), while preserving $M_0$ for $\phi = 0, \pi$.

\section{Magic radii sequence}
\label{MagicRad}

Once the spectrum is known, we can analyze the structure of the electron
ground state of the AA-QD. For definiteness, we examine the dot with
parameter $\phi = \pi$, see Fig.~\ref{fig::spectrum}(a). Let us start,
e.g., with $R=2$. Due to
symmetry~(\ref{eq::spectral_symm}),
we expect that, for the undoped dot, all electron states with negative
eigenenergies are occupied, while all states with positive energies are
empty. The highest occupied electron level
$E_{0;-1}^{+1}$
is separated by a gap
$\Delta_{\rm ex} = E_{0;0}^{-1} - E_{0;-1}^{+1} \approx 0.4 t_0$
from the lowest empty level
$E_{0;0}^{-1}$.
If an electron is promoted from the level
$E_{0;-1}^{+1}$ to $E_{0;0}^{-1}$,
the first excited state of the dot is formed and the excitation energy is
$\Delta_{\rm ex}$.

However, when $R$ grows, the excitation energy
$\Delta_{\rm ex}$
decreases and vanishes at
$R_1 \approx 2.405$
[the first root of the Bessel function
$J_0(z)$].
Here the energies of the ground state and the first excited state coincide.
In other words, the ground state is degenerate with respect to the layer
parity index $c$. The degeneracy equals to two (per valley and per spin
projection). Such a radius will be referred to as ``the first magic
radius''.

The second magic radius is reached at
$R_2 =j_{1;1}\approx 3.8$
[the first non-zero root of
$J_1(z)$],
where
$E_{1;-1}^{+1} (R) = E_{1;0}^{-1} (R)$.
Since at
$\phi = \pi$
all eigenvalues (except those corresponding to
$\mu = 0$)
are doubly degenerate, the extra ground state degeneracy at the second
magic radius is four. Next two magic values are
$R_3 \approx5.1$
($\mu=\pm2$,
$n=0$
and
$n=-1$,
degeneracy is four) and
$R_4 \approx5.5$
($\mu=0$,
$n=1$
and
$n=-2$,
degeneracy is two). More generally, for $\phi = 0$ and $\phi = \pi$ the
magic radius satisfies
\begin{eqnarray}
\label{eq::magic_m_is_zero}
R_i = j_{\mu; n}\,,
\end{eqnarray}
if the dot is undoped.

The latter formula can be generalized to take into account (i)~finite
doping of the dot and (ii)~more general boundary condition, $\phi \ne
0,\,\pi$. To discuss~(i), let us imagine that four electrons (one electron
per valley per spin projection) are removed from the dot. In such a
situation the first magic radius is $R_1 \approx 3.1$, where $E_{0;0}^{-1}$ crosses
$E_{1;-1}^{+1}$, see Fig.~\ref{fig::spectrum}(a). At this radius the ground
state degeneracy of the doped dot increases: at
$R < R_1 \approx 3.1$,
eight electrons completely fill two
$E^{+1}_{\pm 1;-1}$
states in both valleys, while at the magic radius six degenerate orbitals
(two
$E^{+1}_{\pm 1;-1}$
and one
$E^{-1}_{0;0}$
per valley) become available for them. This reasoning further implies
that any crossing point in
Fig.~\ref{fig::spectrum}(a)
represents a magic
radius for a certain doping of the dot. More formally, if radius $R$
satisfies $E_{\mu; n}^{+1} (R) = E_{\mu';n'}^{-1} (R)$ for some $n$, $n'$, $\mu$, and $\mu'$, then one can tune the dot charge to make $R$ a degeneracy point.

As for~(ii), to explore the effects of $\phi \ne 0, \pi$, let us examine
Figs.~\ref{fig::spectrum}(b). For a generic $\phi$ the spectrum loses the
electron-hole symmetry. Yet, as before, the ground state degeneracy
increases at the crossing points of different states, allowing us to define
the magic radii sequence: $R_i$ is a magic radius if
$E_{\mu;n}^{-1} (R_i) = E_{\mu';n'}^{+1} (R_i)$
for some $\mu$, $\mu'$,
$n \geq 0$,
$n' < 0$.

The symmetry between positive and negative parts of the spectrum is
restored at
$\phi = \pi/2$,
see
Fig.~\ref{fig::spectrum}(c).
For this value of $\phi$, the system acquires an additional effective
symmetry. Indeed, one can check that the transformation
\begin{eqnarray}
a_i \rightarrow (-1)^i b_i^*,
\quad
b_i \rightarrow (-1)^i a_i^*,
\end{eqnarray}
inverts the sign of $\varepsilon$ in
Eq.~(\ref{eq::Dirac_BI_layer}),
while preserving the compliance with the boundary condition (the latter
follows from the relation
$M_0 = \sigma_x M_0^* \sigma_x$
valid for
$\phi = \pi/2$).
Note also that, while the eigenvalues spectra are symmetric for
$\phi = \pi/2$
and
$\phi = 0, \pi$,
the magic radii sequence for
$\phi = \pi/2$
differs from that for
$\phi = 0, \pi$.

\section{Numerical results}
\label{Numerical}

\begin{figure}[t]
\includegraphics[width=0.95\columnwidth]{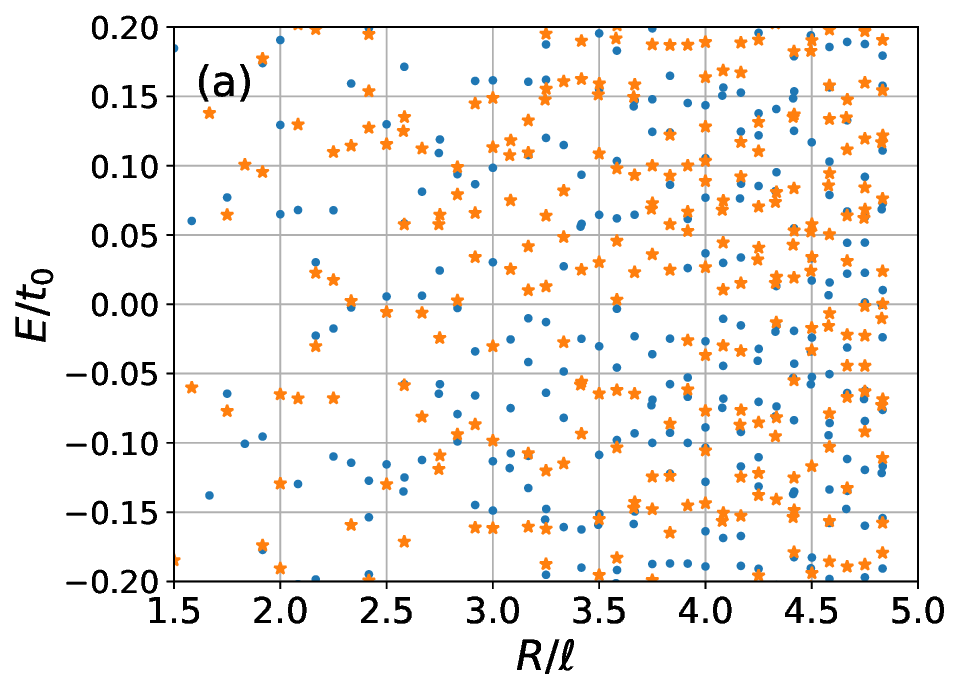}
\includegraphics[width=0.95\columnwidth]{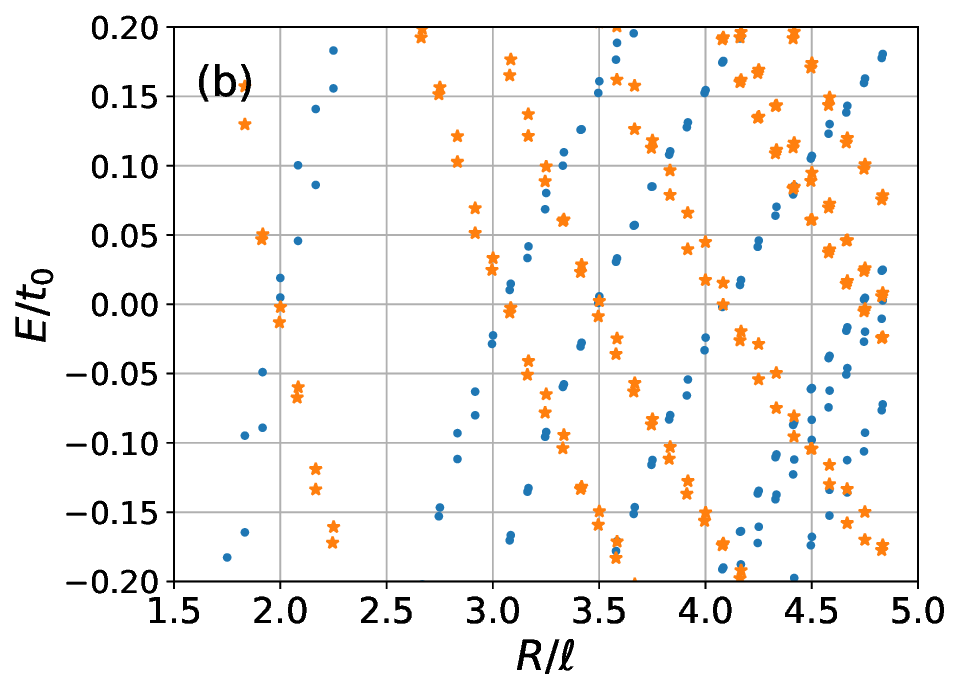}\\
\caption{Numerically calculated spectra of the AA-QD, for different
boundary conditions. Blue dots (yellow stars) show the energies for
layer-symmetric (layer-antisymmetric) states. Similar to
Fig.~\ref{fig::spectrum}(a-c),
when a yellow star comes close or coincides with a blue dot, we have a
magic radius of the AA-QD.
(a)~The spectrum of the AA-QD with ``sharp'' termination. Edge diffusive
scattering randomizes the spectrum, making the magic radii sequence
irregular.
(b)~The same as above, but for a AA-QD with ``soft'' edge, for different
dot radii $R$. Note that the potential
$V_{i\alpha} (\bm{\rho})$
violates electron-hole symmetry of the model [see
Eq.~(\ref{eq::sublattice_symm})],
which leads to the spectrum being weakly asymmetric with respect to
$E \rightarrow -E$
symmetry transformation.
}
\label{fig::spec_numer_main}
\end{figure}

To check and generalize our analytical results, we numerically diagonalize the tight-binding Hamiltonian defined on a circular cluster of radius $R$ [schematically plotted in Fig.~\ref{fig::edge_a}(b)]. The corresponding equations for electron wave functions read
\begin{eqnarray}
\label{eq::TB_SCHR}
&&\varepsilon \Psi_{i \alpha} ({\bf \bm{\rho}})
=
V_{i \alpha} ({\bf \bm{\rho}}) \Psi_{i \alpha} ({\bf \bm{\rho}})
+ t_0 \Psi_{\bar{i} \alpha} ({\bf \bm{\rho}})
\\
\nonumber
&& - t \left[
	\Psi_{i \bar{\alpha}} ({\bf \bm{\rho}}) +
	\Psi_{i \bar{\alpha}} ({\bf \bm{\rho}} + \alpha {\bf a}_1 ) +
	\Psi_{i \bar{\alpha}} ({\bf \bm{\rho}} + \alpha {\bf a}_2 )
	\right],
\end{eqnarray}
where $\Psi_{i \alpha} ({\bf \bm{\rho}})$ represents the wave function for an electron at unit cell ${\bf \bm{\rho}}$, sublattice $\alpha$, in layer $i$. Vectors ${\bf a}_{1,2} = a_0(3, \pm \sqrt{3})/2$ are elementary translations for the graphene lattice. Notation $\bar \alpha$ denotes `not~$\alpha$'. The value of $\alpha$ is $\alpha=1$ ($\alpha=-1$) for sublattice $A$ (sublattice $B$). We assume that the on-site potential $V_{i \alpha} ({\bf \bm{\rho}})$ is layer-independent, $V_{i \alpha} = V_\alpha$, and Eq.~(\ref{eq::TB_SCHR}) splits into two equations describing layer-symmetric and layer-antisymmetric wave functions $\Psi_{i \alpha} \pm \Psi_{\bar{i} \alpha}$.

For a boundary condition to
Eq.~(\ref{eq::TB_SCHR}),
we demand that the electron wave function vanishes outside the dot.
Specifically, for a dot of radius $R$ centered at
${\bf \bm{\rho}}={\bf \bm{\rho}}_0$,
it is required that
$\Psi_{i \alpha } (\bm{\rho}) = 0$
when
$| \bm{\rho}_\alpha - \bm{\rho}_0 | > R$,
where
$\bm{\rho}_\alpha = \bm{\rho} + (\alpha - 1) (\bm{a}_1 + \bm{a}_2)/6$
is the radius-vector for an atom on sublattice $\alpha$ within
$\bm{\rho}$'s
unit cell.

If
$V_{i \alpha} ({\bf \bm{\rho}}) \equiv 0$
everywhere within the AA-QD, we have an AA-QD with a ``sharp'' edge, whose
spectrum is shown in
Fig.~\ref{fig::spec_numer_main}(a).
Such a termination type introduces strong diffusive scattering into the
system due to edge irregularities, see
Fig.~\ref{fig::edge_a}(b).
As a result, the energy spectrum is significantly randomized.

Finite on-site potential at the dot's periphery
$V_{i \alpha} ({\bf \bm{\rho}}) = (-1)^\alpha V_0 (\bm{\rho}_\alpha)$
opens a local gap
$\sim |V_0|$
near the
boundary~\cite{sublatt_symm_gap1, sublatt_symm_gap2}.
The gap pushes electrons away from the dot's boundary, reducing the effect
of the diffusive scattering and making the edge effectively ``soft''. For
numerical calculations, we set
\begin{eqnarray}
\label{potential}
V_0 (\bm{\rho}_\alpha)
=
\overline{V}
\begin{cases}
	0, & \text{if } |\bm{\rho}_\alpha| < R - \delta R, \\
	\frac{|\bm{\rho}_\alpha| - R + \delta R}{\delta R}, &
		\text{if } R - \delta R < |\bm{\rho}_\alpha| < R,
\end{cases}
\end{eqnarray}
where the parameters
$\delta R$
and
$\overline{V}$
are chosen to be
$\delta R = 10 a_0$
and
$\overline{V} = 5$\,eV.
\begin{table}[t]
\centering
\begin{tabular}{|c|c|c|c|}
\hline\hline
$R_i$ & Numerical, & Numerical, & Analytical model, \\ [0.5ex]
& geometric & adjusted & $\phi = \pi/2$ \\ [0.5ex]
\hline
$R_1$ & 2.0 & 1.4 & 1.4 \\
$R_2$ & 3.1 & 2.5 & 2.6 \\
$R_3$ & 3.5 & 2.9 & 3.1 \\
$R_4$ & 4.1 & 3.5 & 3.8 \\ [1ex]
\hline
\end{tabular}
\caption{Four magic radii
$R_1$,..., $R_4$
for circular AA-QDs. All values are dimensionless, in units of $l$. The
second column represents the magic radii for the dots with ``soft'' edge,
as extracted from
Fig.~\ref{fig::spec_numer_main}\,(b).
Since for such an edge the geometric radius of the dot exceeds the radius
of the area accessible to the electrons, we introduce the adjusted radii
which are
0.6$l$
less than the corresponding geometric value, see the third column. The
fourth column shows the radii for our analytical model with the boundary
condition parameter
$\phi = \pi/2$,
per
Fig.~\ref{fig::spectrum}\,(c).
We see a remarkable agreement between the adjusted numerical values and
analytical model values.
\label{tab::radii}
}
\end{table}
The spectra for dots with ``soft'' edges are plotted in
Fig.~\ref{fig::spec_numer_main}\,(b).
We see similar trends as in our analytical model [compare
Fig.~\ref{fig::spec_numer_main}(b) and Fig.~\ref{fig::spectrum}]: one can
discern layer-symmetric (layer-antisymmetric) states, whose energy
increases (decreases) when $R$ grows. Moreover, the spectrum demonstrates
double degeneracy, similar to the degeneracy in our analytical model. The
magic radii sequence can be easily obtained from the results shown in
Fig.~\ref{fig::spec_numer_main}(b).
The first four magic
radii are shown in
Table~\ref{tab::radii}.

Comparing the numerically calculated radii against the results of our
analytical model, it is necessary to keep in mind that the geometric radius
$R$ of an AA-QD with the ``soft'' edge surpasses the radius of the
area accessible to low-energy electrons. Indeed, the potential
$V_0 (\bm{\rho}_\alpha)$
pushes electrons away from dot periphery, turning the ring
$R - \delta R < |\bm{\rho}_\alpha| < R$
into a classically forbidden area. There, a wave function decays as
$\sim \exp \left( - \hbar^{-1} \int dR' |p | \right)$,
where
$|p| = \sqrt{V^2 - \varepsilon^2}/v_{\rm F}  \approx V /v_{\rm F}$ is a
semiclassical momentum of a Dirac-type quantum particle inside the
forbidden ring. Evaluating the integral (it runs from the classical turning
point
$\sim R-\delta R$
to the position
$|\bm{\rho}|$
inside the forbidden area), we extract the characteristic length scale
$l_d \sim \sqrt{\hbar v_{\rm F} \delta R/ \overline{V}} \approx 3 a_0$
over which the wave function decays. Thus, the area accessible for the
electrons inside the dot is limited by the radius
$R - \delta R + 3 a_0 = R - 7 a_0$
for our parameters choice. This means that the geometric radius must be
reduced by
$7a_0$,
or, equivalently, by
$\sim 0.6 l$,
to account for the inaccessible dot's periphery. As for the analytical
model, it is necessary to remember that the magic sequence is sensitive to
the boundary condition parameter $\phi$. We find that
$\phi=\pi/2$
magic radii sequence compares quite well against the (adjusted) magic radii
obtained numerically, see
Table~\ref{tab::radii}.

For ``sharp'' termination, the diffusive scattering makes the spectrum
irregular, see
Fig.~\ref{fig::spec_numer_main}(a).
In this case, the (near) coincidence of the layer-symmetric and
layer-antisymmetric eigenenergies for certain values of $R$ indicate that
this $R$ is a magic radius for a specific doping value.

The numerical results confirm the existence of the magic radii sequence in
the AA-QD. Moreover, this feature of the dot is very stable with respect to
variation of the boundary condition and is a general property that follows
from the layer-symmetry of the AA-based systems.

\section{Discussion}
\label{Discussion}

\subsection{Magic size of the triangular AA-QD}

\begin{figure}[t]
\includegraphics[width=0.95\columnwidth]{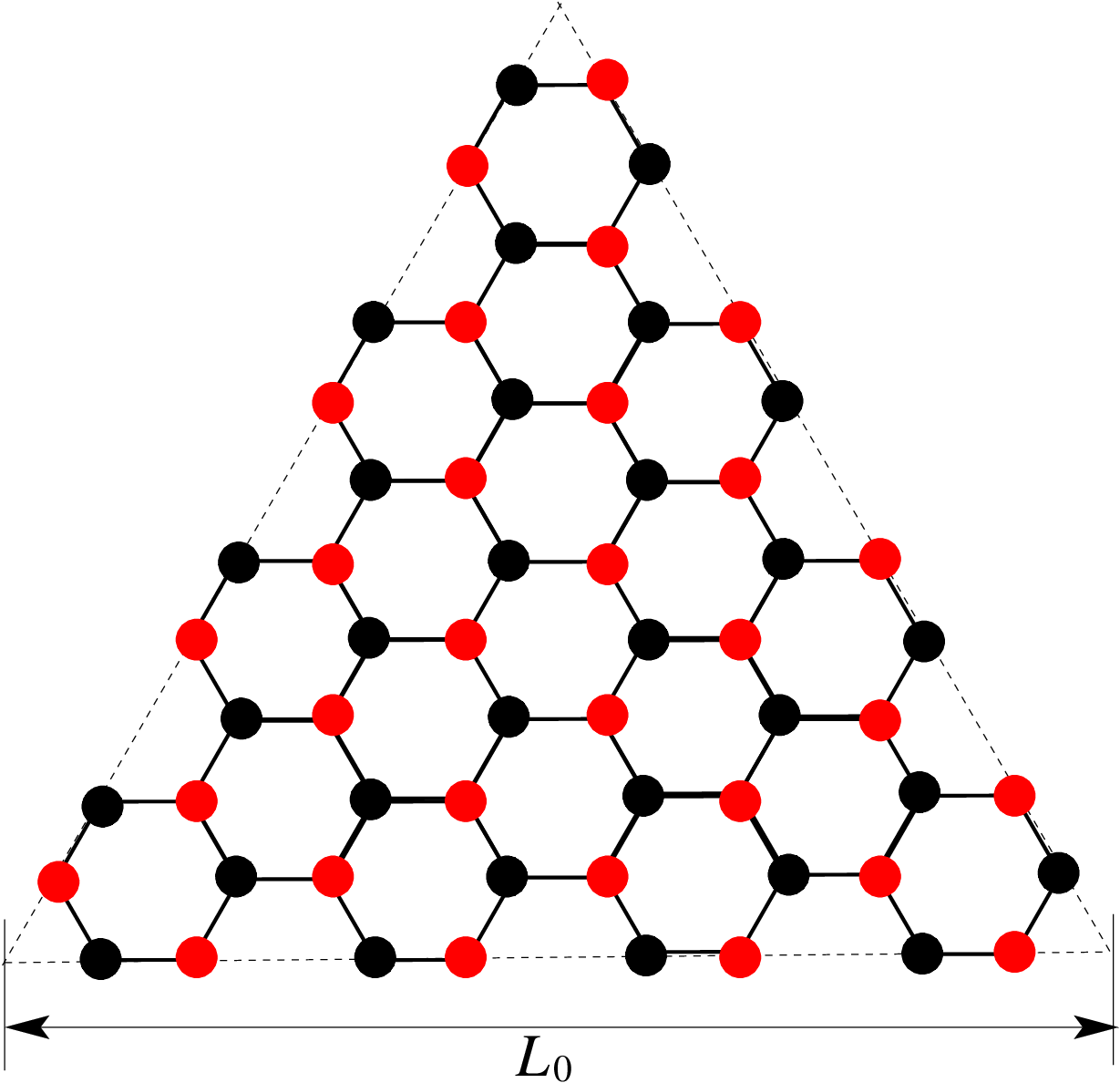}
\caption{The triangular AA-QD, with the armchair-type edges, top view (lower
layer is not visible). Red (black) circles represent the carbon atoms on
$A$ sublattice ($B$ sublattice). The lateral size of the dot is
$L_0$.
\label{fig::triang_AA}
}
\end{figure}

We would like to stress that the degeneracy enhancement at magic values of
$R$ is a direct consequence of the peculiar structure of the AA-BLG
spectrum, and has no immediate connection to the quantum dot shape. To
illustrate this point, we discuss the case of the triangular AA-QD with the
edges of the armchair type, see
Fig.~\ref{fig::triang_AA}.

Triangular single-layer quantum dots with the armchair edges have been
investigated in
Ref.~\onlinecite{triag_dot2010theor}.
There, the analytical expression for the single-electron spectrum has been
derived [see Eq.(81) of
Ref.~\onlinecite{triag_dot2010theor}].
In terms of our notation it reads
$\varepsilon^{\rm SLG}_{m, n, \pm}
\approx
\pm \frac{4 \pi \hbar v_{\rm F}}{3 L_0} \sqrt{n^2 + m^2 - nm}$,
where
$L_0$
is the lateral size of the dot, and
$m \geq 1$
and
$n \geq 1$
are integers.

The dimensionless spectrum of the triangular AA-QD is
\begin{eqnarray}
E^{\rm triang}_{m, n, c, \pm}
\approx
c \pm \frac{4 \pi l }{3 L_0} \sqrt{n^2 + m^2 - nm},
\end{eqnarray}
where $c$, as above, is the layer-parity eigenvalue. The magic value of
$L_0$
is determined by the condition
$E^{\rm triang}_{m, n, +1, -}
=
E^{\rm triang}_{m, n, -1, +}$.
This generates a sequence of magic sizes (in units of $l$) parametrized by
two integers
\begin{eqnarray}
L_{m, n} = \frac{4 \pi }{3 } \sqrt{n^2 + m^2 - nm}.
\end{eqnarray}
Of course, the size
$L_0$
being a discrete quantity changing in multiples of
$3 a_0$
cannot exactly satisfy the latter equation. However, due to smallness of
$3 a_0$
relative to
$4 \pi l/3 \sim 50 a_0$,
a very accurate tuning can be achieved.

Substituting specific values for $n$ and $m$, we calculate
$L_{1,1} \approx 4.19$,
$L_{1,2} \approx 7.26$,
$L_{2,2} \approx 8.38$,
$L_{1,3} \approx 11.08$,
as the first four members of the magic sequence.

\subsection{Comparison with other graphene quantum dots}

We have seen above that an AA-QD has an interesting property: it
demonstrates a sequence of radii at which its ground state experiences an
extended degeneracy. This feature persists regardless of the doping and
boundary condition at the dot boundary and relates with the specific
symmetry of the AA-QD wave functions:
the highest occupied state
is layer-symmetric ($c=+1$) state if $R$ is
somewhat smaller than the nearest magic value $R_m$, while, when $R$
slightly exceeds $R_m$, the highest occupied state is layer-antisymmetric
($c=-1$),
see
Figs.~\ref{fig::spectrum}
and~\ref{fig::spec_numer_main}.
The system has extra degeneracy at
$R=R_m$
when the highest occupied state changes its layer parity. A quantum dot
made from the SLG does not possess this property, neither does a dot from
the AB-BLG. Thus, the existence of the sequence of the magic radii is a
unique feature of the AA-QD.

To illustrate this, let us consider the QD cut out from the SLG (SLG-QD).
The spectrum of the SLG-QD can be obtained directly from the results for
the AA-QD by assigning
$t_0=0$.
In so doing, we derive equation for the energy spectrum in the form
\begin{equation}
\label{eq^SLG_QD}
J_\mu\!\left(\frac{\varepsilon R}{\hbar v_\textrm{F}}\right)+\cot\!\left(\frac{\phi}{2}\right)J_{\mu+1}\!\left(\frac{\varepsilon R}{\hbar v_\textrm{F}}\right)=0\,.
\end{equation}
Thus, the energy spectrum of the SLG-QD scales as
$\varepsilon\propto\pm1/R$,
which is a set of hyperbolas with no magic radii (that is, no crossing
points and extra degeneracy).

\begin{figure}[t]
\includegraphics[width=0.95\columnwidth]{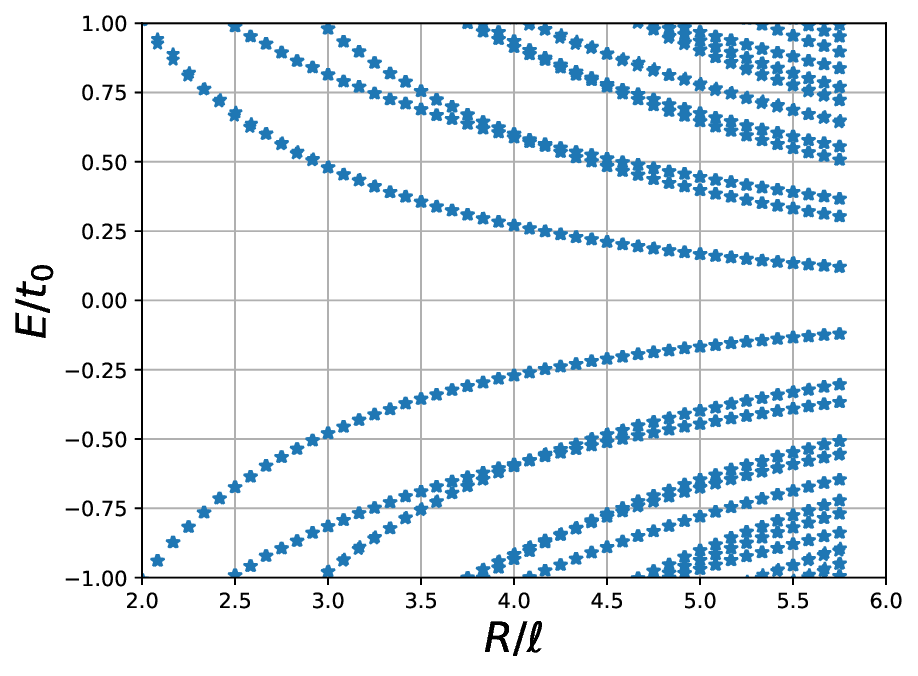}
\caption{Numerically calculated spectra of the AB-QD, with ``soft'' edges.
For any $R$, the spectrum of the undoped dot possesses a finite gap
separating the ground state from excited states.
\label{fig::ab_spec}
}
\end{figure}
For the case of the AB-QD we obtain a similar result. We observe that the
operator $\hat{C}$ does not commute with the Hamiltonian of the AB-BLG: to
map one layer of the AB-BLG to another we should not only apply a mirror
transformation but also perform a shift of the
layers~\cite{ourBLGreview2016}.
As result, there is no quantum number responsible for the extra degeneracy.
A direct calculations show that the eigenenergies of the AB-QD scale as
$\varepsilon\propto\pm 1/R^{2}$.
Numerically calculated low-energy spectra for AB-QD of various radii are
plotted in
Fig.~\ref{fig::ab_spec}.
To soften the effects of the edge irregularities, the on-site potential of
Eq.~(\ref{potential})
is introduced in both layers. The data clearly indicates that for an undoped
AB-QD there is no magic radius, and finite gap separates the ground state
and excited states.

\subsection{Magic radii and magic twist angles}

Discovery of an unusual
superconductivity~\cite{twist_exp_sc2018}
and Mott
transitions~\cite{twist_exp_insul2018}
in the tBLG with the magic twist angle excited keen interest and a flurry
of works on this system. However, a theoretical treatment of the problem
encounters a significant difficulty. Indeed, the first magic twist angle
$\theta$ is about $1^\circ$. Therefore, the tBLG superlattice unit cell
includes several thousands of carbon atoms in non-equivalent positions. The
analytical or numerical study of such a system is a highly non-trivial
task. We believe that our results may be important for the theory of the
twisted bilayer graphene as they can form a basis for development a simple
model of the tBLG with magic twist angles.

To make the connection between the AA-QD and the tBLG, let us consider the
following line of reasoning. The tBLG can be treated as a periodic
structure of AA and AB regions surrounded by areas with some intermediate
stacking types. It is known that the low-energy electron states are
localized within the AA~regions of the tBLG m{\'o}ire
cell~\cite{ourBLGreview2016,PhysRevB.86.155449,PhysRevLett.108.216802,
PhysRevB.92.081406}.
Consequently, one can foresee that the corresponding wave functions may be
described by the formalism developed in this paper. For large-size
m{\'o}ire cell at
$\theta\sim 1^\circ$,
it is also natural to expect, that boundary conditions for a wave function
in the AA region does not give rise to either diffusive or inter-valley
scattering. Then, we can use
Eqs.~(\ref{eq::Dirac1})
and~(\ref{eq::Dirac2})
with boundary
condition~(\ref{eq::dot_BC})
to describe the low-energy states in the tBLG. The hybridization between
m{\'o}ire cells broadens the AA-localized single-electron states into eight
coherent bands (one band per Dirac cone valley, per spin, per layer
parity). These bands are separated from other electron states by a gap of
the order 0.1~eV. This establishes the desired mapping between the spectrum
of the ``magic" AA-QD and the low-energy band structure of the tBLG at the
magic twist
angles~\cite{PhysRevB.86.155449,PhysRevLett.108.216802,twist2015,flatBandViz}.
When the system departs from a ``magic state'' (either magic radius or magic
twist angle), the band arrangement drastically changes: the degeneracy
disappear and the bandwidth increases by order of magnitude. This fact has
also a simple explanation within the proposed framework. Indeed, we can
expect that the band width is the smallest when the bands degeneracy is
maximal. This happens when the AA~region size $R_{\rm AA}$ is close to a
magic radius. Since $R_{\rm AA}$ is proportional to $1/\theta$, the latter
condition can be used to link the magic angle and the magic radius of the
AA region. A characteristic m{\'o}ire size can be estimated as
$L\approx a_0\sqrt3/\theta$
and for the first magic angle we have
$L_1\approx100\,a_0$.
As for
$R_{AA}$,
it is naturally smaller than
$L_1$.
It can be estimated as follows. It is
known~\cite{NanoLettTB,Peeters}
that at small twist angles the supercell of the twisted bilayer graphene
can be considered as consisting of regions with AA, AB, and BA stackings of
the same area. The area of the superlattice cell is
$S_{sc}=\sqrt{3}L_1^2/2$. Thus, we have the estimate:
$R_{AA}\sim\sqrt{S_{sc}/3\pi}\approx0.3L_1\approx30\,a_0$.
According to the results shown in
Fig.~\ref{fig::spec_numer_main},
depending on the boundary conditions, for the first magic radius we have
$17\,a_0 \lesssim R_1 \lesssim 24\,a_0$,
which is consistent with our estimate for
$R_{AA}$
for the tBLG at the first magic angle.

In this paper we focus on the peculiarities of the single-electron spectrum
of the AA-QD. The next step in the study of this structure is to take into
account the electron-electron interaction and to analyze many-body effects.
We believe that the many-body physics of the AA-QD is as rich as the
physics of AA-BLG. Moreover, the features of the single-electron spectrum
of the AA-QD near the magic radii give rise to arising interesting
collective phenomena. We also believe that the study of these phenomena
allows us to understand some important effects in the magic angle tBLG.

\section{Conclusions}
\label{Conclusions}

In conclusions, we have studied the electronic properties of  a circular
quantum dot made from AA bilayer graphene. We observe a discrete set of
``magic'' radii, where the ground state is degenerate with respect to the
layer parity. There is an analogy between ``magic angles" of the twisted
bilayer graphene and ``magic radii" of the AA bilayer graphene quantum dot.
The existence of ``magic radii'' is unique for the AA structures and is
related to the layer-symmetry of the AA graphene bilayer. The analogy between ``magic angles" and ``magic radii" can be helpful for the description of the electronic properties of the twisted bilayer graphene.

\section*{Acknowledgments}
The numerical calculations and data analysis were funded by the RSF grant
No.~22-22-00464,
\url{https://rscf.ru/en/project/22-22-00464/}.
%


\begin{thebibliography}{34}
\expandafter\ifx\csname natexlab\endcsname\relax\def\natexlab#1{#1}\fi
\expandafter\ifx\csname bibnamefont\endcsname\relax
  \def\bibnamefont#1{#1}\fi
\expandafter\ifx\csname bibfnamefont\endcsname\relax
  \def\bibfnamefont#1{#1}\fi
\expandafter\ifx\csname citenamefont\endcsname\relax
  \def\citenamefont#1{#1}\fi

\bibitem[{\citenamefont{Rozhkov et~al.}(2016)\citenamefont{Rozhkov, Sboychakov,
  Rakhmanov, and Nori}}]{ourBLGreview2016}
\bibinfo{author}{\bibfnamefont{A.~V.} \bibnamefont{Rozhkov}},
  \bibinfo{author}{\bibfnamefont{A.~O.} \bibnamefont{Sboychakov}},
  \bibinfo{author}{\bibfnamefont{A.~L.} \bibnamefont{Rakhmanov}},
  \bibnamefont{and} \bibinfo{author}{\bibfnamefont{F.}~\bibnamefont{Nori}},
  {``}\bibinfo{title}{Electronic properties of graphene-based bilayer
  systems},{''} \bibinfo{journal}{Phys. Reports}
  \textbf{\bibinfo{volume}{648}}, \bibinfo{pages}{1} (\bibinfo{year}{2016}).

\bibitem[{\citenamefont{Roy et~al.}(1998)\citenamefont{Roy, Kallinger, and
  Sattler}}]{aa_graphit_roy1998}
\bibinfo{author}{\bibfnamefont{H.-V.} \bibnamefont{Roy}},
  \bibinfo{author}{\bibfnamefont{C.}~\bibnamefont{Kallinger}},
  \bibnamefont{and} \bibinfo{author}{\bibfnamefont{K.}~\bibnamefont{Sattler}},
  {``}\bibinfo{title}{Study of single and multiple foldings of graphitic
  sheets},{''} \bibinfo{journal}{Surf. Sci.} \textbf{\bibinfo{volume}{407}},
  \bibinfo{pages}{1 } (\bibinfo{year}{1998}).

\bibitem[{\citenamefont{Lee et~al.}(2008)\citenamefont{Lee, Lee, Ahn, Kim,
  Wilson, and John}}]{aa_graphite_lee2008}
\bibinfo{author}{\bibfnamefont{J.-K.} \bibnamefont{Lee}},
  \bibinfo{author}{\bibfnamefont{S.-C.} \bibnamefont{Lee}},
  \bibinfo{author}{\bibfnamefont{J.-P.} \bibnamefont{Ahn}},
  \bibinfo{author}{\bibfnamefont{S.-C.} \bibnamefont{Kim}},
  \bibinfo{author}{\bibfnamefont{J.~I.~B.} \bibnamefont{Wilson}},
  \bibnamefont{and} \bibinfo{author}{\bibfnamefont{P.}~\bibnamefont{John}},
  {``}\bibinfo{title}{The growth of AA graphite on (111) diamond},{''}
  \bibinfo{journal}{J. Chem. Phys.} \textbf{\bibinfo{volume}{129}},
  \bibinfo{pages}{234709} (\bibinfo{year}{2008}).

\bibitem[{\citenamefont{Liu et~al.}(2009)\citenamefont{Liu, Suenaga, Harris,
  and Iijima}}]{liu_aa_exp2009}
\bibinfo{author}{\bibfnamefont{Z.}~\bibnamefont{Liu}},
  \bibinfo{author}{\bibfnamefont{K.}~\bibnamefont{Suenaga}},
  \bibinfo{author}{\bibfnamefont{P.~J.~F.} \bibnamefont{Harris}},
  \bibnamefont{and} \bibinfo{author}{\bibfnamefont{S.}~\bibnamefont{Iijima}},
  {``}\bibinfo{title}{Open and Closed Edges of Graphene Layers},{''}
  \bibinfo{journal}{Phys. Rev. Lett.} \textbf{\bibinfo{volume}{102}},
  \bibinfo{pages}{015501} (\bibinfo{year}{2009}).

\bibitem[{\citenamefont{Borysiuk et~al.}(2011)\citenamefont{Borysiuk, Soltys,
  and Piechota}}]{borysiuk2011_aa}
\bibinfo{author}{\bibfnamefont{J.}~\bibnamefont{Borysiuk}},
  \bibinfo{author}{\bibfnamefont{J.}~\bibnamefont{Soltys}}, \bibnamefont{and}
  \bibinfo{author}{\bibfnamefont{J.}~\bibnamefont{Piechota}},
  {``}\bibinfo{title}{Stacking sequence dependence of graphene layers on SiC
  (0001) - Experimental and theoretical investigation},{''}
  \bibinfo{journal}{J. Appl. Phys.} \textbf{\bibinfo{volume}{109}},
  \bibinfo{eid}{093523} (\bibinfo{year}{2011}).

\bibitem[{\citenamefont{Charlier et~al.}(1991)\citenamefont{Charlier,
  Michenaud, Gonze, and Vigneron}}]{aa_tb_charlier1991}
\bibinfo{author}{\bibfnamefont{J.-C.} \bibnamefont{Charlier}},
  \bibinfo{author}{\bibfnamefont{J.-P.} \bibnamefont{Michenaud}},
  \bibinfo{author}{\bibfnamefont{X.}~\bibnamefont{Gonze}}, \bibnamefont{and}
  \bibinfo{author}{\bibfnamefont{J.-P.} \bibnamefont{Vigneron}},
  {``}\bibinfo{title}{Tight-binding model for the electronic properties of
  simple hexagonal graphite},{''} \bibinfo{journal}{Phys. Rev. B}
  \textbf{\bibinfo{volume}{44}}, \bibinfo{pages}{13237} (\bibinfo{year}{1991}).

\bibitem[{\citenamefont{Charlier et~al.}(1992)\citenamefont{Charlier,
  Michenaud, and Gonze}}]{aa_abinit_charlier1992}
\bibinfo{author}{\bibfnamefont{J.-C.} \bibnamefont{Charlier}},
  \bibinfo{author}{\bibfnamefont{J.-P.} \bibnamefont{Michenaud}},
  \bibnamefont{and} \bibinfo{author}{\bibfnamefont{X.}~\bibnamefont{Gonze}},
  {``}\bibinfo{title}{First-principles study of the electronic properties of
  simple hexagonal graphite},{''} \bibinfo{journal}{Phys. Rev. B}
  \textbf{\bibinfo{volume}{46}}, \bibinfo{pages}{4531} (\bibinfo{year}{1992}).

\bibitem[{\citenamefont{Charlier et~al.}(1994)\citenamefont{Charlier, Gonze,
  and Michenaud}}]{aa_graphite_charlier1994}
\bibinfo{author}{\bibfnamefont{J.-C.} \bibnamefont{Charlier}},
  \bibinfo{author}{\bibfnamefont{X.}~\bibnamefont{Gonze}}, \bibnamefont{and}
  \bibinfo{author}{\bibfnamefont{J.-P.} \bibnamefont{Michenaud}},
  {``}\bibinfo{title}{First-principles study of the stacking effect on the
  electronic properties of graphite(s)},{''} \bibinfo{journal}{Carbon}
  \textbf{\bibinfo{volume}{32}}, \bibinfo{pages}{289 } (\bibinfo{year}{1994}).

\bibitem[{\citenamefont{Rakhmanov et~al.}(2012)\citenamefont{Rakhmanov,
  Rozhkov, Sboychakov, and Nori}}]{aa_graph2012_prl}
\bibinfo{author}{\bibfnamefont{A.~L.} \bibnamefont{Rakhmanov}},
  \bibinfo{author}{\bibfnamefont{A.~V.} \bibnamefont{Rozhkov}},
  \bibinfo{author}{\bibfnamefont{A.~O.} \bibnamefont{Sboychakov}},
  \bibnamefont{and} \bibinfo{author}{\bibfnamefont{F.}~\bibnamefont{Nori}},
  {``}\bibinfo{title}{Instabilities of the {$AA$}-Stacked Graphene
  Bilayer},{''} \bibinfo{journal}{Phys. Rev. Lett.}
  \textbf{\bibinfo{volume}{109}}, \bibinfo{pages}{206801}
  (\bibinfo{year}{2012}).

\bibitem[{\citenamefont{Sboychakov
  et~al.}(2013{\natexlab{a}})\citenamefont{Sboychakov, Rakhmanov, Rozhkov, and
  Nori}}]{aa_graph2013}
\bibinfo{author}{\bibfnamefont{A.~O.} \bibnamefont{Sboychakov}},
  \bibinfo{author}{\bibfnamefont{A.~L.} \bibnamefont{Rakhmanov}},
  \bibinfo{author}{\bibfnamefont{A.~V.} \bibnamefont{Rozhkov}},
  \bibnamefont{and} \bibinfo{author}{\bibfnamefont{F.}~\bibnamefont{Nori}},
  {``}\bibinfo{title}{Metal-insulator transition and phase separation in doped
  {AA}-stacked graphene bilayer},{''} \bibinfo{journal}{Phys. Rev. B}
  \textbf{\bibinfo{volume}{87}}, \bibinfo{pages}{121401}
  (\bibinfo{year}{2013}{\natexlab{a}}).

\bibitem[{\citenamefont{Sboychakov
  et~al.}(2013{\natexlab{b}})\citenamefont{Sboychakov, Rozhkov, Rakhmanov, and
  Nori}}]{graph_phasep2013}
\bibinfo{author}{\bibfnamefont{A.~O.} \bibnamefont{Sboychakov}},
  \bibinfo{author}{\bibfnamefont{A.~V.} \bibnamefont{Rozhkov}},
  \bibinfo{author}{\bibfnamefont{A.~L.} \bibnamefont{Rakhmanov}},
  \bibnamefont{and} \bibinfo{author}{\bibfnamefont{F.}~\bibnamefont{Nori}},
  {``}\bibinfo{title}{Antiferromagnetic states and phase separation in doped
  {AA}-stacked graphene bilayers},{''} \bibinfo{journal}{Phys. Rev. B}
  \textbf{\bibinfo{volume}{88}}, \bibinfo{pages}{045409}
  (\bibinfo{year}{2013}{\natexlab{b}}).

\bibitem[{\citenamefont{Akzyanov et~al.}(2014)\citenamefont{Akzyanov,
  Sboychakov, Rozhkov, Rakhmanov, and Nori}}]{aa_graph2014}
\bibinfo{author}{\bibfnamefont{R.~S.} \bibnamefont{Akzyanov}},
  \bibinfo{author}{\bibfnamefont{A.~O.} \bibnamefont{Sboychakov}},
  \bibinfo{author}{\bibfnamefont{A.~V.} \bibnamefont{Rozhkov}},
  \bibinfo{author}{\bibfnamefont{A.~L.} \bibnamefont{Rakhmanov}},
  \bibnamefont{and} \bibinfo{author}{\bibfnamefont{F.}~\bibnamefont{Nori}},
  {``}\bibinfo{title}{{AA}-stacked bilayer graphene in an applied electric
  field: Tunable antiferromagnetism and coexisting exciton order
  parameter},{''} \bibinfo{journal}{Phys. Rev. B}
  \textbf{\bibinfo{volume}{90}}, \bibinfo{pages}{155415}
  (\bibinfo{year}{2014}).

\bibitem[{\citenamefont{Mosoyan et~al.}(2018)\citenamefont{Mosoyan, Rozhkov,
  Sboychakov, and Rakhmanov}}]{graphIt_aa}
\bibinfo{author}{\bibfnamefont{K.~S.} \bibnamefont{Mosoyan}},
  \bibinfo{author}{\bibfnamefont{A.~V.} \bibnamefont{Rozhkov}},
  \bibinfo{author}{\bibfnamefont{A.~O.} \bibnamefont{Sboychakov}},
  \bibnamefont{and} \bibinfo{author}{\bibfnamefont{A.~L.}
  \bibnamefont{Rakhmanov}}, {``}\bibinfo{title}{Spin-density wave state in
  simple hexagonal graphite},{''} \bibinfo{journal}{Phys. Rev. B}
  \textbf{\bibinfo{volume}{97}}, \bibinfo{pages}{075131}
  (\bibinfo{year}{2018}).

\bibitem[{\citenamefont{Brey and Fertig}(2013)}]{brey_fertig_aa2013}
\bibinfo{author}{\bibfnamefont{L.}~\bibnamefont{Brey}} \bibnamefont{and}
  \bibinfo{author}{\bibfnamefont{H.~A.} \bibnamefont{Fertig}},
  {``}\bibinfo{title}{Gapped phase in $AA$-stacked bilayer graphene},{''}
  \bibinfo{journal}{Phys. Rev. B} \textbf{\bibinfo{volume}{87}},
  \bibinfo{pages}{115411} (\bibinfo{year}{2013}).

\bibitem[{\citenamefont{Sboychakov et~al.}(2021)\citenamefont{Sboychakov,
  Rakhmanov, Rozhkov, and Nori}}]{fracmet2021prblett}
\bibinfo{author}{\bibfnamefont{A.~O.} \bibnamefont{Sboychakov}},
  \bibinfo{author}{\bibfnamefont{A.~L.} \bibnamefont{Rakhmanov}},
  \bibinfo{author}{\bibfnamefont{A.~V.} \bibnamefont{Rozhkov}},
  \bibnamefont{and} \bibinfo{author}{\bibfnamefont{F.}~\bibnamefont{Nori}},
  {``}\bibinfo{title}{Bilayer graphene can become a fractional metal},{''}
  \bibinfo{journal}{Phys. Rev. B} \textbf{\bibinfo{volume}{103}},
  \bibinfo{pages}{L081106} (\bibinfo{year}{2021}).

\bibitem[{\citenamefont{Chiu et~al.}(2010)\citenamefont{Chiu, Lee, Chen, Shyu,
  and Lin}}]{optical_theor_aa2010}
\bibinfo{author}{\bibfnamefont{C.~W.} \bibnamefont{Chiu}},
  \bibinfo{author}{\bibfnamefont{S.~H.} \bibnamefont{Lee}},
  \bibinfo{author}{\bibfnamefont{S.~C.} \bibnamefont{Chen}},
  \bibinfo{author}{\bibfnamefont{F.~L.} \bibnamefont{Shyu}}, \bibnamefont{and}
  \bibinfo{author}{\bibfnamefont{M.~F.} \bibnamefont{Lin}},
  {``}\bibinfo{title}{Absorption spectra of AA-stacked graphite},{''}
  \bibinfo{journal}{New J. Phys.} \textbf{\bibinfo{volume}{12}},
  \bibinfo{pages}{083060} (\bibinfo{year}{2010}).

\bibitem[{\citenamefont{Alam et~al.}(2011)\citenamefont{Alam, Lin, and
  Saito}}]{bi3}
\bibinfo{author}{\bibfnamefont{M.~S.} \bibnamefont{Alam}},
  \bibinfo{author}{\bibfnamefont{J.}~\bibnamefont{Lin}}, \bibnamefont{and}
  \bibinfo{author}{\bibfnamefont{M.}~\bibnamefont{Saito}},
  {``}\bibinfo{title}{First-Principles Calculation of the Interlayer Distance
  of the Two-Layer Graphene},{''} \bibinfo{journal}{Jpn. J. Appl. Phys.}
  \textbf{\bibinfo{volume}{50}}, \bibinfo{pages}{080213}
  (\bibinfo{year}{2011}).

\bibitem[{\citenamefont{Chang}(2012)}]{optical_theor_aa_ab2012}
\bibinfo{author}{\bibfnamefont{C.~P.} \bibnamefont{Chang}},
  {``}\bibinfo{title}{Analytic model of energy spectrum and absorption spectra
  of bilayer graphene},{''} \bibinfo{journal}{J. Appl. Phys.}
  \textbf{\bibinfo{volume}{111}}, \bibinfo{eid}{103714} (\bibinfo{year}{2012}).

\bibitem[{\citenamefont{Apinyan and Kope{\'{c}}}(2021)}]{aa_graph_afm2021theor}
\bibinfo{author}{\bibfnamefont{V.}~\bibnamefont{Apinyan}} \bibnamefont{and}
  \bibinfo{author}{\bibfnamefont{T.~K.} \bibnamefont{Kope{\'{c}}}},
  {``}\bibinfo{title}{Antiferromagnetic ordering and excitonic pairing in
  AA-stacked bilayer graphene},{''} \bibinfo{journal}{Phys. Rev. B}
  \textbf{\bibinfo{volume}{104}}, \bibinfo{pages}{075426}
  (\bibinfo{year}{2021}).

\bibitem[{\citenamefont{Rozhkov et~al.}(2017)\citenamefont{Rozhkov, Rakhmanov,
  Sboychakov, Kugel, and Nori}}]{spin_valley2017prl}
\bibinfo{author}{\bibfnamefont{A.~V.} \bibnamefont{Rozhkov}},
  \bibinfo{author}{\bibfnamefont{A.~L.} \bibnamefont{Rakhmanov}},
  \bibinfo{author}{\bibfnamefont{A.~O.} \bibnamefont{Sboychakov}},
  \bibinfo{author}{\bibfnamefont{K.~I.} \bibnamefont{Kugel}}, \bibnamefont{and}
  \bibinfo{author}{\bibfnamefont{F.}~\bibnamefont{Nori}},
  {``}\bibinfo{title}{Spin-Valley Half-Metal as a Prospective Material for Spin
  Valleytronics},{''} \bibinfo{journal}{Phys. Rev. Lett.}
  \textbf{\bibinfo{volume}{119}}, \bibinfo{pages}{107601}
  (\bibinfo{year}{2017}).

\bibitem[{\citenamefont{Akhmerov and Beenakker}(2008)}]{akhmerov}
\bibinfo{author}{\bibfnamefont{A.~R.} \bibnamefont{Akhmerov}} \bibnamefont{and}
  \bibinfo{author}{\bibfnamefont{C.~W.~J.} \bibnamefont{Beenakker}},
  {``}\bibinfo{title}{Boundary conditions for Dirac fermions on a terminated
  honeycomb lattice},{''} \bibinfo{journal}{Phys. Rev. B}
  \textbf{\bibinfo{volume}{77}}, \bibinfo{pages}{085423}
  (\bibinfo{year}{2008}).

\bibitem[{\citenamefont{Sanderson et~al.}(2013)\citenamefont{Sanderson, Ang,
  and Zhang}}]{PhysRevB.88.245404}
\bibinfo{author}{\bibfnamefont{M.}~\bibnamefont{Sanderson}},
  \bibinfo{author}{\bibfnamefont{Y.~S.} \bibnamefont{Ang}}, \bibnamefont{and}
  \bibinfo{author}{\bibfnamefont{C.}~\bibnamefont{Zhang}},
  {``}\bibinfo{title}{Klein tunneling and cone transport in AA-stacked bilayer
  graphene},{''} \bibinfo{journal}{Phys. Rev. B} \textbf{\bibinfo{volume}{88}},
  \bibinfo{pages}{245404} (\bibinfo{year}{2013}).

\bibitem[{\citenamefont{Giovannetti et~al.}(2007)\citenamefont{Giovannetti,
  Khomyakov, Brocks, Kelly, and van~den Brink}}]{sublatt_symm_gap1}
\bibinfo{author}{\bibfnamefont{G.}~\bibnamefont{Giovannetti}},
  \bibinfo{author}{\bibfnamefont{P.~A.} \bibnamefont{Khomyakov}},
  \bibinfo{author}{\bibfnamefont{G.}~\bibnamefont{Brocks}},
  \bibinfo{author}{\bibfnamefont{P.~J.} \bibnamefont{Kelly}}, \bibnamefont{and}
  \bibinfo{author}{\bibfnamefont{J.}~\bibnamefont{van~den Brink}},
  {``}\bibinfo{title}{Substrate-induced band gap in graphene on hexagonal boron
  nitride: Ab initio density functional calculations},{''}
  \bibinfo{journal}{Phys. Rev. B} \textbf{\bibinfo{volume}{76}},
  \bibinfo{pages}{073103} (\bibinfo{year}{2007}).

\bibitem[{\citenamefont{Gusynin et~al.}(2007)\citenamefont{Gusynin, Sharapov,
  and Carbotte}}]{sublatt_symm_gap2}
\bibinfo{author}{\bibfnamefont{V.}~\bibnamefont{Gusynin}},
  \bibinfo{author}{\bibfnamefont{S.}~\bibnamefont{Sharapov}}, \bibnamefont{and}
  \bibinfo{author}{\bibfnamefont{J.}~\bibnamefont{Carbotte}},
  {``}\bibinfo{title}{AC conductivity of graphene: from tight-binding model to
  2+ 1-dimensional quantum electrodynamics},{''}
  \bibinfo{journal}{International Journal of Modern Physics B}
  \textbf{\bibinfo{volume}{21}}, \bibinfo{pages}{4611} (\bibinfo{year}{2007}).

\bibitem[{\citenamefont{Rozhkov and Nori}(2010)}]{triag_dot2010theor}
\bibinfo{author}{\bibfnamefont{A.~V.} \bibnamefont{Rozhkov}} \bibnamefont{and}
  \bibinfo{author}{\bibfnamefont{F.}~\bibnamefont{Nori}},
  {``}\bibinfo{title}{Exact wave functions for an electron on a graphene
  triangular quantum dot},{''} \bibinfo{journal}{Phys. Rev. B}
  \textbf{\bibinfo{volume}{81}}, \bibinfo{pages}{155401}
  (\bibinfo{year}{2010}).

\bibitem[{\citenamefont{Cao et~al.}(2018{\natexlab{a}})\citenamefont{Cao,
  Fatemi, Fang, Watanabe, Taniguchi, Kaxiras, and
  Jarillo-Herrero}}]{twist_exp_sc2018}
\bibinfo{author}{\bibfnamefont{Y.}~\bibnamefont{Cao}},
  \bibinfo{author}{\bibfnamefont{V.}~\bibnamefont{Fatemi}},
  \bibinfo{author}{\bibfnamefont{S.}~\bibnamefont{Fang}},
  \bibinfo{author}{\bibfnamefont{K.}~\bibnamefont{Watanabe}},
  \bibinfo{author}{\bibfnamefont{T.}~\bibnamefont{Taniguchi}},
  \bibinfo{author}{\bibfnamefont{E.}~\bibnamefont{Kaxiras}}, \bibnamefont{and}
  \bibinfo{author}{\bibfnamefont{P.}~\bibnamefont{Jarillo-Herrero}},
  {``}\bibinfo{title}{Unconventional superconductivity in magic-angle graphene
  superlattices},{''} \bibinfo{journal}{Nature} \textbf{\bibinfo{volume}{556}},
  \bibinfo{pages}{43} (\bibinfo{year}{2018}{\natexlab{a}}).

\bibitem[{\citenamefont{Cao et~al.}(2018{\natexlab{b}})\citenamefont{Cao,
  Fatemi, Demir, Fang, Tomarken, Luo, Sanchez-Yamagishi, Watanabe, Taniguchi,
  Kaxiras et~al.}}]{twist_exp_insul2018}
\bibinfo{author}{\bibfnamefont{Y.}~\bibnamefont{Cao}},
  \bibinfo{author}{\bibfnamefont{V.}~\bibnamefont{Fatemi}},
  \bibinfo{author}{\bibfnamefont{A.}~\bibnamefont{Demir}},
  \bibinfo{author}{\bibfnamefont{S.}~\bibnamefont{Fang}},
  \bibinfo{author}{\bibfnamefont{S.~L.} \bibnamefont{Tomarken}},
  \bibinfo{author}{\bibfnamefont{J.~Y.} \bibnamefont{Luo}},
  \bibinfo{author}{\bibfnamefont{J.~D.} \bibnamefont{Sanchez-Yamagishi}},
  \bibinfo{author}{\bibfnamefont{K.}~\bibnamefont{Watanabe}},
  \bibinfo{author}{\bibfnamefont{T.}~\bibnamefont{Taniguchi}},
  \bibinfo{author}{\bibfnamefont{E.}~\bibnamefont{Kaxiras}},
  \bibnamefont{et~al.}, {``}\bibinfo{title}{Correlated insulator behaviour at
  half-filling in magic-angle graphene superlattices},{''}
  \bibinfo{journal}{Nature} \textbf{\bibinfo{volume}{556}}, \bibinfo{pages}{80}
  (\bibinfo{year}{2018}{\natexlab{b}}).

\bibitem[{\citenamefont{Lopes~dos Santos et~al.}(2012)\citenamefont{Lopes~dos
  Santos, Peres, and Castro~Neto}}]{PhysRevB.86.155449}
\bibinfo{author}{\bibfnamefont{J.~M.~B.} \bibnamefont{Lopes~dos Santos}},
  \bibinfo{author}{\bibfnamefont{N.~M.~R.} \bibnamefont{Peres}},
  \bibnamefont{and} \bibinfo{author}{\bibfnamefont{A.~H.}
  \bibnamefont{Castro~Neto}}, {``}\bibinfo{title}{Continuum model of the
  twisted graphene bilayer},{''} \bibinfo{journal}{Phys. Rev. B}
  \textbf{\bibinfo{volume}{86}}, \bibinfo{pages}{155449}
  (\bibinfo{year}{2012}).

\bibitem[{\citenamefont{San-Jose et~al.}(2012)\citenamefont{San-Jose,
  Gonz\'alez, and Guinea}}]{PhysRevLett.108.216802}
\bibinfo{author}{\bibfnamefont{P.}~\bibnamefont{San-Jose}},
  \bibinfo{author}{\bibfnamefont{J.}~\bibnamefont{Gonz\'alez}},
  \bibnamefont{and} \bibinfo{author}{\bibfnamefont{F.}~\bibnamefont{Guinea}},
  {``}\bibinfo{title}{Non-Abelian Gauge Potentials in Graphene Bilayers},{''}
  \bibinfo{journal}{Phys. Rev. Lett.} \textbf{\bibinfo{volume}{108}},
  \bibinfo{pages}{216802} (\bibinfo{year}{2012}).

\bibitem[{\citenamefont{Yin et~al.}(2015)\citenamefont{Yin, Qiao, Zuo, Li, and
  He}}]{PhysRevB.92.081406}
\bibinfo{author}{\bibfnamefont{L.-J.} \bibnamefont{Yin}},
  \bibinfo{author}{\bibfnamefont{J.-B.} \bibnamefont{Qiao}},
  \bibinfo{author}{\bibfnamefont{W.-J.} \bibnamefont{Zuo}},
  \bibinfo{author}{\bibfnamefont{W.-T.} \bibnamefont{Li}}, \bibnamefont{and}
  \bibinfo{author}{\bibfnamefont{L.}~\bibnamefont{He}},
  {``}\bibinfo{title}{Experimental evidence for non-Abelian gauge potentials in
  twisted graphene bilayers},{''} \bibinfo{journal}{Phys. Rev. B}
  \textbf{\bibinfo{volume}{92}}, \bibinfo{pages}{081406}
  (\bibinfo{year}{2015}).

\bibitem[{\citenamefont{Sboychakov et~al.}(2015)\citenamefont{Sboychakov,
  Rakhmanov, Rozhkov, and Nori}}]{twist2015}
\bibinfo{author}{\bibfnamefont{A.~O.} \bibnamefont{Sboychakov}},
  \bibinfo{author}{\bibfnamefont{A.~L.} \bibnamefont{Rakhmanov}},
  \bibinfo{author}{\bibfnamefont{A.~V.} \bibnamefont{Rozhkov}},
  \bibnamefont{and} \bibinfo{author}{\bibfnamefont{F.}~\bibnamefont{Nori}},
  {``}\bibinfo{title}{Electronic spectrum of twisted bilayer graphene},{''}
  \bibinfo{journal}{Phys. Rev. B} \textbf{\bibinfo{volume}{92}},
  \bibinfo{pages}{075402} (\bibinfo{year}{2015}).

\bibitem[{\citenamefont{Utama et~al.}(2021)\citenamefont{Utama, Koch, Lee,
  Leconte, Li, Zhao, Jiang, Zhu, Watanabe, Taniguchi et~al.}}]{flatBandViz}
\bibinfo{author}{\bibfnamefont{M.~I.~B.} \bibnamefont{Utama}},
  \bibinfo{author}{\bibfnamefont{R.~J.} \bibnamefont{Koch}},
  \bibinfo{author}{\bibfnamefont{K.}~\bibnamefont{Lee}},
  \bibinfo{author}{\bibfnamefont{N.}~\bibnamefont{Leconte}},
  \bibinfo{author}{\bibfnamefont{H.}~\bibnamefont{Li}},
  \bibinfo{author}{\bibfnamefont{S.}~\bibnamefont{Zhao}},
  \bibinfo{author}{\bibfnamefont{L.}~\bibnamefont{Jiang}},
  \bibinfo{author}{\bibfnamefont{J.}~\bibnamefont{Zhu}},
  \bibinfo{author}{\bibfnamefont{K.}~\bibnamefont{Watanabe}},
  \bibinfo{author}{\bibfnamefont{T.}~\bibnamefont{Taniguchi}},
  \bibnamefont{et~al.}, {``}\bibinfo{title}{Visualization of the flat
  electronic band in twisted bilayer graphene near the magic angle twist},{''}
  \bibinfo{journal}{Nat. Phys.} \textbf{\bibinfo{volume}{17}},
  \bibinfo{pages}{184} (\bibinfo{year}{2021}).

\bibitem[{\citenamefont{Trambly~de Laissardi\`{e}re
  et~al.}(2010)\citenamefont{Trambly~de Laissardi\`{e}re, Mayou, and
  Magaud}}]{NanoLettTB}
\bibinfo{author}{\bibfnamefont{G.}~\bibnamefont{Trambly~de Laissardi\`{e}re}},
  \bibinfo{author}{\bibfnamefont{D.}~\bibnamefont{Mayou}}, \bibnamefont{and}
  \bibinfo{author}{\bibfnamefont{L.}~\bibnamefont{Magaud}},
  {``}\bibinfo{title}{Localization of Dirac Electrons in Rotated Graphene
  Bilayers},{''} \bibinfo{journal}{Nano Lett.} \textbf{\bibinfo{volume}{10}},
  \bibinfo{pages}{804} (\bibinfo{year}{2010}).

\bibitem[{\citenamefont{An{\dj}elkovi{\'{c}}
  et~al.}(2018)\citenamefont{An{\dj}elkovi{\'{c}}, Covaci, and
  Peeters}}]{Peeters}
\bibinfo{author}{\bibfnamefont{M.}~\bibnamefont{An{\dj}elkovi{\'{c}}}},
  \bibinfo{author}{\bibfnamefont{L.}~\bibnamefont{Covaci}}, \bibnamefont{and}
  \bibinfo{author}{\bibfnamefont{F.~M.} \bibnamefont{Peeters}},
  {``}\bibinfo{title}{DC conductivity of twisted bilayer graphene:
  Angle-dependent transport properties and effects of disorder},{''}
  \bibinfo{journal}{Phys. Rev. Materials} \textbf{\bibinfo{volume}{2}},
  \bibinfo{pages}{034004} (\bibinfo{year}{2018}).

\end{thebibliography}

\end{document}